\documentclass[12pt]{article}
\usepackage{graphicx}
\usepackage{amsmath}

\begin{document}
\begin{center}\Large {\bf
    Weak Lie symmetry and extended Lie algebra}\normalsize \\\end{center}\begin{center} Hubert  
  Goenner\\ Institute for Theoretical Physics\\ Friedrich-Hund-Platz 1
  \\ University of Goettingen\end{center} 

\vspace{4cm} 

Abstract:\footnote{A summary of this article has been presented at the ``90th 
Encounter between Mathematicians and Theoretical Physicists'' at the Institut de
 Recherche Math\'ematique Avanc\'ee (University of Strasbourg and CNRS),
    September 20-22, 2012.}\\

The concept of weak Lie motion (weak Lie symmetry) is introduced through 
${\cal{L}}_{\xi}{\cal{L}}_{\xi}g_{ab}=0,$ (${\cal{L}}_{\xi}{\cal{L}}_{\xi}f=0$). 
Applications are given which exhibit a reduction of the usual
symmetry, e.g., in the case of the the rotation group. In this
context, a particular generalization of Lie algebras is found
(``extended Lie algebras'') which turns out to be an involutive
distribution or a simple example for a tangent Lie
algebroid. Riemannian and Lorentz metrics can be introduced on such an 
algebroid through an extended Cartan-Killing form. Transformation
groups from non-relativistic mechanics and quantum mechanics lead to
such tangent Lie algebroids and to Lorentz geometries constructed on
them (1-dimensional gravitational fields).  

\pagebreak
\section{Introduction}
In 1872 Felix Klein formulated his Erlangen program as such: ``A
manifold is given and with it a group of transformations. [..] Develop
the theory of invariants with regard to this group''
(\cite{Klein1927}, p. 28). According to him, Sophus Lie accepted this
program and spread it among his students.\footnote{The history of the
  ``Erlanger Programm'' is much more complicated, though,
  cf. \cite{Rowe89}, \cite{Hawk1989}. For the importance of Klein’s
  ideas for physics cf. \cite{Kastrup87}.}    

In the first part of what follows, in the spirit of Klein, several new
concepts will be introduced and investigated: {\em weak (Lie) motions}
(cf. section \ref{section:weakmotion}) and {\em groups of extended
  motions} (cf. section \ref{section:extmotion}). The latter concept
is related to a suggested widening of the physicists' concept of a Lie
algebra to particular tangent Lie algebroids  ({\em   extended Lie
  algebras}). Some of the corresponding finite transformations forming
groups are presented: they are no longer Lie groups. I will also
propose an extension of the Cartan-Killing form which up to now
seemingly has not been studied. Its definition allows the introduction
of Riemannian and Lorentz metrics on the sections of a subbundle of
the tangent bundle. The mathematical literature for algebroids and
groupoids  (eg., \cite{Mack2005}), has lead to a few formal
applications to Lagrangian mechanics \cite{Liber1996},
\cite{Martin2001}, \cite{Grabo2006}. The particular tangent Lie
algebroids presented here are an example for such structures much closer to
physics than the examples usually given by mathematicians.  

\section{Lie-dragging}
\label{section:Liedrag}
\subsection{Preliminaries}
\label{subsection: prelim}
In metric geometry, the concept of symmetry may be expressed by an
isometry of the metrical tensor $g_{ab}$ of such a space. This means
that this tensor field remains unchanged along the flow of a vector
field $X$. An expression for this demand may be formulated by help of
the Lie derivative defined for tangent vector fields
$X:=\xi^{a} \frac{\partial}{\partial x^{a}}, Y:=\eta^{a}
\frac{\partial}{\partial x^{a}}$ by:\begin{equation}  {\cal{L}}_{X} Y 
  = [X, Y], \label{Liederi}\end{equation} where $[.~,~.] $  
 denotes the Lie-bracket $[A, B] =  AB-BA$. If (\ref{Liederi}) is expressed 
 by the components $\xi^{a}, \eta^{a}$ of the tangent vectors $X, Y$, then 
 \begin{equation} {\cal{L}}_{\xi}
   \eta^{a} = \eta^{a}_{~, c} \xi^{c} - \eta^{c} \xi^{a}_{~,
     c}~, \label{Liedericomp}\end{equation} where $\eta^{a}_{~,
   c}=\frac{\partial \eta^{a}}{\partial x^{c}}$. If ${\cal{L}}_{X} Y
 =0$, the vector field $X$ is called a symmetry of the vector field
 $Y$.\footnote{Such symmetries play an important role for the
   integration of differential equation, cf. \cite{Wino97}.} The
 Leibniz rule holds for the Lie derivative.\footnote{Latin indices
   from the beginning ($a, b, c,..$) and end of the alphabet ($r,s,
   t,..$) run from 1 to n or 0 to n-1 where n is the dimension of the
   space considered. Indices from the middle ($i, j, k, l, ..$) may take
  other values. The summation convention is used except when indicated 
  otherwise.} From (\ref{Liederi}) we have  \begin{equation}
  {\cal{L}}_{Z} {\cal{L}}_{X} Y = [Z,
  [X,Y]], \label{Liederi2},\end{equation} and with help of the Jacobi 
 identity:  \begin{equation} {\cal{L}}_{Z} {\cal{L}}_{X} Y +  {\cal{L}}_{Y}
   {\cal{L}}_{Z} X +  {\cal{L}}_{X} {\cal{L}}_{Y} Z= [Z, [X, Y]] +
   [Y, [Z, X]] +  [X, [Y, Z]] = 0. \label{Jaco}\end{equation} From
 (\ref{Jaco}): \begin{equation} {\cal{L}}_{Z} {\cal{L}}_{X} Y -
   {\cal{L}}_{X} {\cal{L}}_{Z} Y= [[X,Z], Y] =  {\cal{L}}_{[X,Z]} Y =
   {\cal{L}}_{{\cal{L}}_{X}Z} Y~. \label{redu}\end{equation}\\ 
For a Lie group, a special subspace of the tangent space is formed by
the infinitesimal generators
$X_{(i)}:=\xi_{(i)}^{a}\frac{\partial}{\partial x^{a}},~ (i, j, l = 1,
2, .., p)$ of a Lie-algebra \begin{equation}  [X_{(i)}, 
  X_{(j})] = c_{ij}^{~~l}  X_{(l)}~, \label{Liealgeb}\end{equation}
with  structure constants\footnote{In current mathematical literature,
the definitition of a Lie algebra is much more general. It is defined
either as a module ${\cal B}(M)$ of the set of all $ C^{\infty}$-vector
fields on a $C^{\infty}$-manifold with a multiplication introduced via the
Lie-bracket, or as a finite-dimensional vector space $V$ over the real
or complex numbers with a bilinear multiplication on it defined by an
anti-commuting bracket $[~ ,~ ]$ satisfying the Jacobi identity
(\ref{Jaco}).} $ c_{ij}^{~~l}$. From (\ref{Liealgeb}) we
obtain: \begin{equation}{\cal{L}}_{X_{i}} {\cal{L}}_{X_{j}} X_{k} =  c_{jk}^{~~l}
  c_{il}^{~~m}X_{m}\label{Liealg2}\end{equation} such that according
to (\ref{Jaco}):\begin{equation} c_{jk}^{~~l} c_{il}^{~~m}
  +c_{ij}^{~~l} c_{kl}^{~~m}+c_{ki}^{~~l} c_{jl}^{~~m}
  =0. \label{Jaco2}\end{equation} A symmetric bilinear form, the
Cartan-Killing form, may be introduced:\begin{equation} \sigma_{ij}:=
  c_{il}^{~~m}c_{jm}^{~~~l}~.\end{equation} If it is nondegenerate, i.e.,
for semisimple Lie groups, $\sigma_{ij}$ can be used as a metric in
group space. 

In section \ref{section:algebext}, we will permit that the structure constants
become directly dependent on the components $\xi^{a}_{~i}$ of the
vector fields $X_{i}(x)$: they will become {\em structure
  functions}.\footnote{The {\em structure constants} in (\ref{Liealgeb}) are 
  brought into the definitions of a Lie algebra presented in the
  previous footnote   by the choice of a basis $\{Y_1, Y_2,.. ,Y_n\}$
  of $V$. The multiplicative action  is  determined for all vectors $
  X, Y$ of $V$ only if all brackets $[X, Y]$ are known. According to
  one author: ``We 'know' them by writing them as linear combinations
  of the $Y_{i}$. The coefficients $   c_{ij}^{~~l}$ in the relations
  $[Y_{i}, Y_{j}]= c_{ij}^{~~l}  X_{l}$ are called structure
  constants'' (\cite{Samel89}, pp. 1, 5). This recipee no longer works for vector
  fields which cannot be generated by linear combinations with
  constant coefficients from a basis. Cf. section \ref{section:algebext}.}\\  

\subsection{Lie-dragging (with examples)}
\label{subsection:Liedragging}
Under ``Lie-dragging'' with regard to an arbitrary $C^{\infty}$ vector
field $X=\xi^{a} \frac{\partial}{\partial x^{a}}$ we understand the
operation of the Lie derivative on any geometric object {\em without}
the simultaneous requirement that the result be zero.\footnote{This use 
of the name ``Lie-dragging'' is different from the one in
\cite{Schutz82}. By (\ref{Liederi}), the Lie-dragging of a vector
field is expressed.} Applied to the metric $g_{ab}$, this
means \begin{equation}{\cal{L}}_{\xi} g_{ab} =
  \gamma_{ab}~, \label{liedrag} \end{equation} where $\gamma_{ab}$ is
a symmetric tensor of any rank between 0 and n (in n-dimensional
space). In the sequel we will be interested in the case $\gamma_{ab}
\neq \lambda g_{ab}$.  

For a tensor field, Lie-dragging neither conserves the rank of the field, 
nor, if it is excerted on a symmetric bilinear form, its
signature. The quest for the conditions that Lie-dragging leads to a
specific rank or specific signature of a tensor field could be among
the first mathematical investigations into the concept (with rank 0 of
$\gamma_{ab}$ being set aside). Also, the vector fields $X$ might be
classified according to whether Lie-dragging with them leads to a
prescribed rank for given metric $g_{ab}$. In any case, not every
arbitrary $\gamma_{ab}$ can be reached by Lie-dragging (cf. Appendix 1).\\
\noindent Equation (\ref{liedrag}) can be read in different ways:\\ 

\noindent A) Given a single vector field (a set of vector fields)
{\em and} an arbitrary metric $g_{ab}$; the set of all possible
bilinear forms $\gamma_{ab}$ is to be determined by a straightforward
calculation. This is an intermediate step for the determination of
weak Lie   motions of $g_{ab}$.\\ 

\noindent B)  Given a single vector field (a set of vector fields)
and a fixed target tensor $\gamma_{ab};$ the metrics $g_{ab}$ which
are Lie-dragged into it are to be determined. This requires solving a system of
1st-order PDEs.\\  

\noindent C) Given both a start metric $g_{ab}$ and a target metric
$\gamma_{ab}$. The task is to determine the vector fields $X$ dragging
the one into the other.\footnote{If we ask for both,
  ${\cal{L}}_{X}g_{ab} = \gamma_{ab}$ and  ${\cal{L}}_{X}\gamma_{ab} =
  g_{ab}$, then we are back to weak homothetic mappings for both $g$
  and $\gamma$. Cf. next section.} \\

For a first example for Lie-dragging in space-time leading to
tensors of lower rank, we look at the Kasner metric: 
\begin{equation} ds^2 = (dx^0)^2 - (x^0)^{2p_1}(dx^1)^2 -
  (x^0)^{2p_2}(dx^2)^2 -(x^0)^{2p_3}(dx^3)^2~, \label{Kasner}\end{equation} an 
 exact solution of Einstein's vacuum field equations if $p_1 + p_2 +
 p_3 = 1 = (p_1)^2+(p_2)^2+(p_3)^2~, p_1, p_2, p_3$
 constants. Lie-dragging  with \begin{center} $X=\delta_0^a \frac{\partial}{\partial x^a}$ \end{center} leads to a bilinear form of rank 3, i.e., after a coordiante change, to the space sections:  \begin{equation} ds^2 =  - (y^0)^{2p_1}(dy^1)^2 - (y^0)^{2p_2}(dy^2)^2 -
  (y^0)^{2p_3}(dy^3)^2. \nonumber \end{equation} Unlike this,
Lie-dragging of (\ref{Kasner}) with \begin{center}$X= f(x^0)\delta_1^a
\frac{\partial}{\partial x^a}$ \end{center} leads to a   tensor of rank 2:
$\gamma_{ab}= 2 \frac{df(x^0)}{dx^0}~ g_{1(a}\delta_{b)}^{~0}$. 

In the second example, a Lie-dragged metric of rank 1 is
prescribed. Let \begin{equation} {\cal{L}}_{\xi} g_{ab}  = X_a X_b
  ~,\end{equation} with the vector field $X$ tangent to a null
geodesic:\begin{equation}(\overset{g}{\nabla}_{b}X_{a}) X^{b}= 0~,
  g_{ab} X^{a}X^{b}= 0\label{examp1} ~. \end{equation} From
the definition of $ {\cal{L}}_{\xi} g_{ab}$ given in (\ref{isomet2})
and (\ref{examp1}), $(X^{s} \xi_s)_{,a}X^a =0 $ follows: $X^{s} \xi_s$
must be constant along the geodesic. (\ref{examp1}) leads to a
restriction on $\xi$ for given null geodesic, or for $X^a$ if the
vector field $\xi$ is given. $X^a$ generates a super-weak motion
(cf. section \ref{section:weakmotion}).\\

 The collineations presented in section \ref{section:motions} are also
examples for Lie-dragging.

\section{Motions and Collineations}
\label{section:motions} On a manifold with differentiable metric
structure, a motion is defined by the vanishing of the Lie-derivative
of the metric with regard to the tangent vector field
$X=\xi^a\frac{\partial}{\partial 
  x^{a}}$: \begin{eqnarray} {\cal{L}}_{X} g(Y,Z) = 0 = X g(Y,Z) +
  g(Z,{\cal{L}}_{X}Y) + g(Y,{\cal{L}}_{X}Z) \nonumber\\ =  X g(Y,Z) +
  g(Z,[X,Y]) + g(Y,[X,Z]), \label{Liecoofree}\end{eqnarray} where $X, Y, 
Z$ are tangent vector fields. In local coordinates, (\ref{Liecoofree})
reads as: \begin{equation} \gamma_{ab}= {\cal{L}}_{\xi} g_{ab}=0= g_{ab,c}~\xi^{c}
  + g_{cb}~\xi^{c}_{~,a} + g_{ac}~\xi^{c}_{~,b}
  ~, \label{isomet}\end{equation} with $g_{ab}= g_{ba}$. The vector
field $\xi$ is named a {\em Killing vector}; its components generate an infinitesimal 
symmetry transformation:\footnote{For mechanical systems in phase space, this 
  infinitesimal symmetry transformation is applied to the generalized
  coordinates and supplemented by an infinitesimal transformation for
  the momenta: $p_a \rightarrow p_{a'}=p_a + \eta_a $ with an additional
  infinitesimal generator $\eta_a$. Cf. \cite{Pang2009}. The authors
  use the name ``weak-Lie'' symmetry for what we would name Lie
  symmetry.} $x^{i} \rightarrow x^{i'}= x^{i} +
\xi^{i}$. (\ref{isomet}) may be expressed in a different form:\footnote{Symmetrization 
brackets are used: $A_{(r}B_{s)}=   \frac{1}{2}(A_rB_s + A_sB_r);~A_{[r}B_{s]}= 
  \frac{1}{2}(A_rB_s - A_sB_r)$.}\begin{equation}{\cal{L}}_{\xi} 
  g_{ab} =  2\overset{g}{\nabla}_{(a}\xi_{b)} = 0 .\label{isomet2}\end{equation}
In (\ref{isomet2}), $\overset{g}{\nabla}$ is the covariant derivative
with respect to the metric $g_{ab}$ (Levi Civita connection), and
$\xi_a =g_{ab}\xi^{b}$. From (\ref{isomet}) we can conclude that
${\cal{L}}_{\xi}ds= 0$ for all $dx^a$, i.e., all distances remain invariant. A 
consequence of (\ref{isomet}) is that the motions $\xi$ form a Lie group and 
the corresponding infinitesimal generators 
$X_{(i)}:=\xi_{(i)}^{\sigma} \frac{\partial}{\partial x^{\sigma}}$ a
Lie algebra (\ref{Liealgeb}) (cf. \cite{Yano57}). \\ 

As an example for a group of motions in 3-dimensional Euclidean space, we start
from a Lie group $G_3$ acting on $V_3$ with finite transformations:
\begin{equation} x^{1'} = x^{1} + c_{1},~ x^{2'} = x^{2} + c_{2}x^{1},~x^{3'} = x^{3}
+ c_{3} ~.\label{fintrans} \end{equation} The corresponding Lie
algebra is (\cite{Petrov1969}, p. 213): \begin{equation} [X_1, X_2]=0,~ [X_1,
  X_3]=0,~[X_2, X_3]=   X_1~. \label{LieG3} \end{equation} 
Lie-dragging with the vector fields
$\xi^a_1=\delta^a_2,~\xi^a_2=\delta^a_3,~\xi^a_3=-\delta^a_1 
+ x^3\xi^a_2 $ gives:\begin{eqnarray}{\cal{L}}_{\xi_1} g_{ab} =
  g_{ab,2}=:  \overset{(1)}{\gamma}_{ab}~,~{\cal{L}}_{\xi_2} g_{ab} =
  g_{ab,3}=:  \overset{(2)}{\gamma}_{ab}~, ~\nonumber\\{\cal{L}}_{\xi_3} g_{ab} =
  -g_{ab,1} + x^3 g_{ab,2} + 2 g_{2(a}\delta_{b)}^3 =:
  \overset{(3)}{\gamma}_{ab}~. \end{eqnarray} All $ \overset{(i)}{\gamma}_{ab}$
 can have full rank. 
The demand $\overset{(i)}{\gamma}_{ab}=0,~ i= 1, 2, 3 ,$ makes this $G_3$ a group of 
motions whence follows:      
\begin{gather*} g_{ab}= \begin{pmatrix} \alpha_{11}^{(0)} &
    \alpha_{12}^{(0)} & P_1 \\ \alpha_{21}^{(0)} & \alpha_{22}^{(0)} &
    P'_1 \\ P_1 & P'_1 & P_2 \end{pmatrix}, \label{G3}\end{gather*} where
$P_1= \alpha_{12}^{(0)} x^{1} +  \alpha_{13}^{(0)}, P'_1 =
\alpha_{22}^{(0)} x^{1} +  \alpha_{23}^{(0)}$ and $P_2
=\alpha_{22}^{(0)} (x^{1})^{2} + 2 \alpha_{23}^{(0)}x^{1} +
\alpha_{33}^{(0)}$ with $\alpha_{33}^{(0)}, \alpha_{1p}^{(0)},
\alpha_{2p}^{(0)}~, (p = 1, 2, 3)$ constants. We will see in section
\ref{G3weak} how the metric looks if the group is demanded to be a complete
set of {\em weak} (Lie) motions.\\ 

Further types of symmetries are defined by the vanishing
of the Lie derivative applied to other geometric objects like connection
(``affine collineations''  ${\cal{L}}_{\xi}~
  \Gamma^{~~~c}_{ab}(g)= 0$, cf. \cite{Maar87}), curvature tensor
(``curvature collineations''  ${\cal{L}}_{\xi}~R^{c}_{~dab}(g)=0$, cf. \cite{KaLeDa1969}), 
Ricci tensor (``Ricci'' or ``contracted curvature collineations'' ${\cal{L}}_{\xi}~R^{c}_{~abc}(g)=0,$
cf. \cite{Collin1970}). Another generalization is the concept of conformal Killing vector, defined 
by: \begin{equation} {\cal{L}}_{\xi} g_{ab} = \lambda (x^1,..x^n) 
  g_{ab} ~, \label{isomet3}~.\end{equation} A subcase are {\em homothetic} 
motions with $\lambda = \lambda_{0} =$ const. Conformal Killing vectors are 
included in what follows. Thus, (\ref{isomet}) and
(\ref{isomet3}) are particular subcases of Lie-dragging: they
constitute a fixed point in the map of symmetric differentiable tensor
fields $g_{ab}$ of full rank defined by Lie-dragging.

\section{Weak Lie motions (weak symmetries)}
\label{section:weakmotion}
In the 80s, a concept of ``p-invariance'' has been
introduced \cite{Papa83}:\begin{equation}
  {\cal{L}}_{\xi}......{\cal{L}}_{\xi}~g_{ab} = 0~, \end{equation} 
with p Lie derivatives, $p>1$, acting on the metric. At the time, for
$p=2$ an application has been given in Einstein-Maxwell theory \cite{Goe84}. In 
the following we will concentrate on this case $p=2$.\\

{\em Definition 1}:\\
An infinitesimal point transformation $x \rightarrow x+\xi$ satisfying
\begin{equation}{\cal{L}}_{\xi}{\cal{L}}_{\xi}g_{ab}=0,~{\cal{L}}_{\xi}g_{ab}\neq
  0, \label{isomet5}\end{equation} generates a ``weak Lie motion''.\\
A coordinate-free formulation of (\ref{isomet5}) is: \begin{equation}
{\cal{L}}_{W}{\cal{L}}_{Z} g(X,Y)= [W, Z] g(X,Y) - g([W,[Z,X]], Y) -
g(X, [Y,[W,Z]]). \nonumber\end{equation}  

\noindent If applied to other geometric objects, we call
(\ref{isomet5}) ``weak symmetry''.\footnote{In the set of solutions of
  (\ref{isomet5}), the isometries (motions) must also occur. We speak
  of {\em genuine} weak Lie motions when motions are to be
  excluded.}We also use the expression {\em weak isometry}. 

\noindent ({\ref{isomet5}) can be read in two ways:\\

 - The metric $g_{ab}$ is given; determine the generator $\xi$ of a
 weak Lie motion;\\ 

 - A vector field or a Lie algebra is given; determine the metric
 $g_{ab}$ which allows these fields as weak Lie motions.\\

As has been pointed out in \cite{Papa83}, a disadvantage of the new concept
is that ${\cal{L}}_{\xi}{\cal{L}}_{\xi}g^{ab} = 0$ does not follow from
${\cal{L}}_{\xi}{\cal{L}}_{\xi}g_{ab} = 0$ for ${\cal{L}}_{\xi}g_{ab}\neq 0$. In
fact:\begin{equation}{\cal{L}}_{\xi}{\cal{L}}_{\xi}g^{ab} =
  -g^{as}g^{bt}{\cal{L}}_{\xi}{\cal{L}}_{\xi}g_{st} +
  2g^{at}g^{bp}g^{sq}({\cal{L}}_{\xi}g_{pq})
  ({\cal{L}}_{\xi}g_{st})~. \label{difsym}\end{equation} Consequently,
in general ${\cal{L}}_{\xi}{\cal{L}}_{\xi}g^{ab} = 0$ and
${\cal{L}}_{\xi}{\cal{L}}_{\xi}g_{ab} = 0$ define slightly different
invariance concepts. If both conditions are imposed,
${\cal{L}}_{\xi}g_{ab} = \Phi (x) k_{a}k_{b}$ with the null vector $
k_{a}~ (g^{rs} k_{r}k_{s}=0)$, and arbitrary scalar function $\Phi$
follows. In this case, we call the weak motion generated by $X =\xi^{a}
\frac{\partial}{\partial x^{a}}$ a {\em super weak motion}. It entails
the existence of a null vector $ k_{a}$ with ${\cal{L}}_{\xi} k^{a} = 
- k^{a}{\cal{L}}_{\xi}(ln \Phi)$.\footnote{In general relativity, $T^{ab}= 
\Phi (x) k_{a}k_{b}$ describes a null-fluid.  What is called here
super-weak motion, would have be named {\em cosymmetric-2-invariance} in
(\cite{Papa83}, p. 138).} In Euclidean space ${\cal{L}}_{\xi}g_{ab} =
0$ results. For $p>2$ the situation would become still more complicated.\\  

\subsection{First examples and generalizations}

\subsubsection{Weak symmetries}
\label{subsubsection:weaksym}
That a weak symmetry can be really weaker than a symmetry is
seen already when the Lie derivative is applied twice to a function
$f(x^1,... x^n)$: \begin{equation}  {\cal{L}}_{X}{\cal{L}}_{X}f =
  {\cal{L}}_{\xi}{\cal{L}}_{\xi}f = XXf \overset{!}{=}
  0~.\label{Lie2}\end{equation} In n-dimensional Euclidean space 
  $R^n$, for a translation in the direction of the k-axis with
$\xi^{i}=\delta^{i}_{(k)}$, we obtain from (\ref{Lie2}) $f= x^k
f_1(x^1,..,x^{k-1}, x^{k+1},..,..x^n) + f_2((x^1,..,x^{k-1},
x^{k+1},..,..x^n)$ in place of $f= f((x^1,..,x^{k-1},
x^{k+1},..,..x^n)$  for ${\cal{L}}_{\xi}f \overset{!}{=} 0$. For the
full translation group of  $R^n$, (\ref{Lie2}) leads to a polynomial of degree $n$ 
in the variables $(x^1,..,x^{n})$ with constant coefficients and linear in each variable 
$(x^1,..,x^{n})$. Thus, for  $n=3$, $f=c_{123} x^1 x^2 x^3 + \Sigma_{r, s=1;r<s,}^{3}c_{rs} 
x^{r} x^{s} +\Sigma_{s=1}^{3}c_{s}x^{s} + c_{0}$ as compared to $f=f_0$ for the 
translation group as a group of motions.\footnote{Note that this result follows only if definition 3 
for a complete set of weak symmetries is applied, cf. next sextion.}\\
For a rotation $R^{i}_{k} = x^{i}\frac{\partial}{\partial
  x^{k}}-x^{k}\frac{\partial}{\partial x^{i}}~ (i, k$ {\em fixed}), a
function satisfying ${\cal{L}}_{\xi}{\cal{L}}_{\xi}f \overset{!}{=}0$
\linebreak is given by  $f= \alpha_1(x^1,..,x^{i-1},
x^{i+1},..x^{k-1}, x^{k+1},..x^n)\times  arctan\frac{x^{i}}{x^{k}}$
\linebreak + $\alpha_2(x^1,..,x^{i-1}, x^{i+1},..x^{k-1},
x^{k+1},..x^n)$, with ${\cal{L}}_{\xi}f= -\alpha_{1} \neq 0 $  for this
rotation. For the full rotation group $SO(3)$ in 3-dimensional space, $f= f (\sqrt{(x^1)^2 + (x^2)^2 +(x^3)^2}~)$ follows: no genuine weak motion is possible in this case.
These examples show that the set of weak-Lie invariant {\em functions} can be larger.\\  

A {\em generalization} of a subgroup of the abelian translation group in an
n-dimensional euclidean space is given by: \begin{equation} x^{1'} = x^{1} +
  G^{1}(x^{k+1},.., x^{n}),..,x^{k'} = x^{k} + G^{k}(x^{k+1},..,x^{n}),
  x^{(k+1)'}=x^{k+1},..,x^{n'}=x^{n}, \label{freefunc}\end{equation}
with arbitrary $C^{\infty}$ functions $G^1, G^2,.., G^k$.
Weak Lie symmetry under this group for the function $f(x^{1},..,x^{n})$ leads to the same result
as for the translation group, although (\ref{freefunc}) no longer is a Lie group

A link between weak Lie symmetry of scalars and weak Lie motions can be found
in conformally flat metrics: $g_{ab}= f(x^1, x^2,.., x^n) \eta_{ab}$ due to 
\begin{equation}  {\cal{L}}_{\xi} {\cal{L}}_{\xi} g_{ab}=
  ({\cal{L}}_{\xi} {\cal{L}}_{\xi} f) \eta_{ab} + 2 {\cal{L}}_{\xi} f~
  {\cal{L}}_{\xi} \eta_{ab} + {\cal{L}}_{\xi} {\cal{L}}_{\xi}
  \eta_{ab} \end{equation} 
In the special case of (\ref{isomet3}) follows: \begin{equation} {\cal{L}}_{\xi}
  {\cal{L}}_{\xi} g_{ab} = (\lambda^2+   \lambda_{,s}\xi^{s}) g_{ab}
  ~,~ {\cal{L}}_{\xi} {\cal{L}}_{\xi}   g^{ab} = (\lambda^2-
  \lambda_{,s}\xi^{s}) g^{ab}   ~. \label{isomet4}\end{equation}
Hence, in this case nothing new is   obtained by letting the Lie-derivative act
twice. The concept of conformal Killing   vector could also
be weakend to {\em  weak conformal Killing vector} by the
demand: \begin{equation}   {\cal{L}}_{\xi}{\cal{L}}_{\xi}g_{ab}=
  \lambda(x^i)g_{ab}~,~{\cal{L}}_{\xi}g_{ab}\neq \mu(x^j) g_{ab}~.\end{equation}

\subsubsection{Weak collineations}
For {\em weak Lie affine collineations}, we find:\begin{eqnarray} {\cal{L}}_{\xi} 
{\cal{L}}_{\xi} \Gamma^{~~~c}_{ab}(g) = \xi^{s}
  \overset{g}{\nabla}_{(a}[{\cal{L}}_{\xi} \Gamma^{~~~c}_{b)s}(g)] +
  [{\cal{L}}_{\xi} \Gamma^{~~~c}_{bs}(g)]\overset{g}{\nabla}_{a}\xi^{s}
  \nonumber\\ + [{\cal{L}}_{\xi} \Gamma^{~~~c}_{as}(g)]
  \overset{g}{\nabla}_{b}\xi^{s} - [{\cal{L}}_{\xi}
  \Gamma^{~~~s}_{ab}(g)]\overset{g}
  {\nabla}_{s}\xi^{c}~.\label{bicol1}\end{eqnarray} 
Insertion of $ {\cal{L}}_{\xi}~  \Gamma^{~~~c}_{ab}(g) =  \overset{g}{\nabla}_{a}
  \overset{g}{\nabla}_{b}\xi^{c}+ R^{c}_{~bda}(g)\xi^{d}$ into (\ref{bicol1}) leads
to:\begin{eqnarray} {\cal{L}}_{\xi} {\cal{L}}_{\xi}
  \Gamma^{~~~c}_{ab}(g) = \xi^{s} \overset{g}{\nabla}_{(a}
  \overset{g}{\nabla}_{b)}\overset{g}{\nabla}_{s}\xi^{c} + \xi^{s}
  [\overset{g}{\nabla}_{(a} R^{c}_{~|ds|b)}]\xi^{d} +  \xi^{s}
  R^{c}_{~|ds|(b} \overset{g}{\nabla}_{a)}\xi^{d}\nonumber\\ + R^{c}_{~dbs}\xi^{d}
  ~\overset{g}{\nabla}_{a}\xi^{s} + R^{c}_{~das}\xi^{d}
  ~\overset{g}{\nabla}_{b}\xi^{s} - R^{s}_{~dab}\xi^{d}
  ~\overset{g}{\nabla}_{s}\xi^{c} +
  \overset{g}{\nabla}_{b}\overset{g}{\nabla}_{s}\xi^{c}\overset{g}{\nabla}_{a}\xi^{
    s} \nonumber \\+
  \overset{g}{\nabla}_{a}\overset{g}{\nabla}_{s}\xi^{c}\overset{g}{\nabla}_{b}\xi^{
    s}
  -\overset{g}{\nabla}_{a}\overset{g}{\nabla}_{b}\xi^{s}\overset{g}{\nabla}_{s}\xi^
 {c}~. \end{eqnarray}   
In Minkowski space, the condition is obtained: \begin{equation}\xi^{s}\partial_{a} \partial_{b} \partial_{s}~
  \xi^{c} + \partial_{b} \partial_{s} \xi^{c}~ \partial_{a}
  \xi^{s} + \partial_{a} \partial_{s} \xi^{c}~ \partial_{b}
  \xi^{s} - \partial_{a} \partial_{b} \xi^{s}~ \partial_{s}
  \xi^{c} = 0~ .   \end{equation} to be satisfied by the
generators of the weak Lie   affine collineation. A particular solution is
given by $ \xi^{c}= \beta^{c}f(\alpha_{rs} x^{r}x^{s})$ with constants
$\alpha_{rs}, \beta^{c}$ and $\beta^{s}\alpha_{sa}= 0$ and
arbitrary $C^{3}$-function $f$.    

If spaces with a Riemannian (Lorentzian) metric are considered, the
following expression for weak affine collineations
obtains: \begin{equation}{\cal{L}}_{\xi}{\cal{L}}_{\xi}\{_{ab}^{c}\}=
  -g^{cp}g^{sq} \gamma_{pq}[\overset{g}{\nabla}_{(a}\gamma_{b)s}
  -\frac{1}{2}\overset{g}{\nabla}_{s}{\cal{L}}_{\xi}\gamma_{ab}] +
  g^{cs} [\overset{g}{\nabla}_{(a}{\cal{L}}_{\xi}\gamma_{b)s} -
  \frac{1}{2}\overset{g}{\nabla}_{s}\gamma_{ab} -
  {\cal{L}}_{\xi}\{_{ab}^{t}\}\gamma_{st}]~, \label{riemcollin} \end{equation}
where $\gamma_{ab}$ was defined in (\ref{liedrag}). The concept of
weak Lie   curvature collineations could also be introduced:~
${\cal{L}}_{\xi}{\cal{L}}_{\xi}R^{c}_{~dab}(g)=0$. This concept leads
to 4th-order PDEs.  
\subsection{Complete sets of weak Lie motions}
\label{subsection:weakogroup}
If $g_{ab}$ allows the maximal group of motions with $(^{n+1}_{~~2})$
parameters, no genuine weak Lie motions do exist. If $g_{ab}$ allows a
r-parameter group of motions, then $(^{n+1}_{~~2}) -r$ 
genuine weak Lie motions may exist. The case of a Lie group with
$(^{n+1}_{~~2}) - 1$ parameters acting as an isometry group cannot occur
in n-dimensional space (Fubini 1903). Hence, in space-time which allows
a 10-parameter group as maximal group, no 9-parameter Lie group
exists. For 4-dimensional Lorentz-space (with signature $\pm 2$), 8-parameter
Lie groups are likewise excluded as isometry groups (Jegorov 1955)
(\cite{Petrov1969}, p. 134).\footnote{This does not hold for Finsler
  geometry by which an 8-parameter Lie groups is admitted. Cf. \cite{Bogo1977},
  \cite{Bogo1994}, \cite{Bogoen1999}} Thus, besides the maximal
group, the largest group of motions in space-time is a 7-parameter
group.\footnote{Petrov's claim that for 4-dimensional Lorentz spaces
  7-parameter Lie groups are excluded, is not correct,
  cf. \cite{Petrov1969}, p. 134), \cite{bible}, p. 122).} In this
case, the largest group of weak Lie motions would then be a
3-parameter Lie group.\\   

According to (\ref{redu}), a consequence for weak motions
is:\footnote{If an extended Lie algebra is used, on the r.h.s. of
  (\ref{Liebrak2}), the term  $2 c_{ji~, (a}^{~~k}
  g_{b)c}\xi_{k}^{c}$ must be added.} \begin{equation}    
({\cal{L}}_{\xi_{i}}{\cal{L}}_{\xi_{j}} -
  {\cal{L}}_{\xi_{j}}{\cal{L}}_{\xi_{i}}) g_{ab}=
  {\cal{L}}_{({\cal{L}}_{\xi_{i}}\xi_{j})} g_{ab}=
  {\cal{L}}_{c_{ji}^{~k}\xi_{k}} g_{ab}
  \\=c_{ji}^{~k}{\cal{L}}_{\xi_{k}}g_{ab}.\label{Liebrak2}\end{equation}   
  
\noindent (\ref{Liebrak2}) provides a hint about how a {\em group of
weak Lie symmetries} is to be defined when a set of vector fields, $\xi,
\eta, \zeta, ..$ has been found satifying (\ref{isomet5}). For genuine weak motions, not 
all of the following equations can be satisfied: 
${\cal{L}}_{\eta}{\cal{L}}_{\xi}g_{ab}= 0,~$
${\cal{L}}_{\xi}{\cal{L}}_{\eta}g_{ab}=
0,~{\cal{L}}_{\eta}{\cal{L}}_{\zeta}g_{ab}= 
0,~{\cal{L}}_{\zeta}{\cal{L}}_{\eta}g_{ab}= 0,~ 
{\cal{L}}_{\zeta}{\cal{L}}_{\xi}g_{ab}= 0,
~{\cal{L}}_{\xi}{\cal{L}}_{\zeta}g_{ab}= 0,~ ... .$ If the r vectors
$\xi_{(k)}, k= 1, 2, .. , r$ are the infinitesimal generators of a Lie,
group, the above demand {\em in general} leads into an impasse:
instead of its intended role as a weak Lie-invariance group, it
reduces to an isometry group. This is due to (\ref{redu}) or
(\ref{Liebrak2}). An exception holds if some of the vector fields
commute. 

Consequently, the following definition may be introduced:\\ 
 
{\em Definition 2} (strong complete set):\\  A Lie algebra presents a
strong complete set of weak Lie symmetries if at least one of the
corresponding Lie algebra elements does not generate a motion ($
{\cal{L}}_{\xi_{(j)}} g_{ab}\neq 0$ for one $(j)$, at least) and the
following $(^{m+1}_{~2})~, m>1$ conditions hold:\begin{equation}
  {\cal{L}}_{\xi_{(i)}}{\cal{L}}_{\xi_{(j)}}   g_{ab}=
  0,~ \end{equation} for $(i)=(j)$  and $(i)<(j),  (i), (j) =  
1, 2, .., m$ or, for  $(i)=(j)$  and $(i)>(j),~ (i), (j) = 1, 2, .., m.$\\  
The remaining ${\cal{L}}_{\xi_{(i)}}{\cal{L}}_{\xi_{(j)}} g_{ab}\neq 0$
for $(i)>(j)~ [(i)<(j)]$ are then determined through (\ref{redu}). In
general, we will demand that none of the vector fields $X_{(i)}$ generate motions.

A less demanding definition would be:\\

{\em Definition 3} (complete set):\\  A  Lie algebra leads to a
complete set of weak Lie-symmetries if each of its infinitesimal operators
$X_{i}=\xi_{(i)}^{~a}\frac{\partial}{\partial x^{a}}$ generates a weak
Lie motion: ${\cal{L}}_{\xi_{(i)}}{\cal{L}}_{\xi_{(i)}} g_{ab} = 
0,~{\cal{L}}_{\xi_{(i)}} g_{ab} \neq 0$ for every $ i = 1, 2,
...,m.$\\ 

In section \ref{subsection:G3weak}, examples will be given showing
that the alternative definitions 2 and 3 for complete sets of weak Lie
symmetries lead to different results. In general, we will prefer
definition 2.\\ 

As will be seen in the next section, a consequence is that if $g(X,Y)$
allows the {\em maximal} group of motions, weak Lie motions for
$g(X,Y)$ do not exist or reduce to conformal motions. As an example:
in 2-dimensional Euclidean space with a 3-parameter maximal group (two
translations and one rotation), no genuine weak (Lie) motion exists. The other
extremal case is the non-existence of genuine weak Lie motions, e.g., for the
rotation group together with definition 2. The Kasner metric
(\ref{Kasner}) which allows three space translations as isometries,
is a candidate for not leading to genuine weak Lie motions.

\section{Weak Lie invariance}
\label{section:weakinvar}
\noindent We now want to determine the metrics allowing a
time translation and the rotation group as weak Lie motions.
The group is chosen such that, as an isometry group, it describes
{\em static, spherically symmetric (s.s.s.) metrics}. Thus we have to
allow for four vector fields $\xi_{(i)}, i=1, 2, 3, 4 $ forming a Lie
algebra with a 2-parameter abelian subalgebra and then drag twice the
arbitrary metric $g_{ab} $. At first, definition 3 is applied and the target
metric $\gamma_{ab}$ calculated. 

\subsection{Weakly static metrics.}
To begin, we demand that only the time translation $T= X_1$ with components 
$ \xi_{(1)}^{s}= \delta_{0}^{s}$ generates a weak motion:
${\cal{L}}_{X_1}{\cal{L}}_{X_1}g_{ab} = 0$. The resulting class of
metrics is: \begin{equation} g_{ab} = x^0 c_{ab}(x^1, x^2,  x^3) +
  d_{ab}(x^1, x^2,  x^3)~,\label{metgstat}\end{equation} 
with arbitrary symmetric tensors $c_{ab}, d_{ab}$. The class remains
invariant with regard to linear transformations in time $x^0
\rightarrow \alpha(x^1, x^2,  x^3) x^0 +\beta(x^1, x^2,  x^3);~
\alpha, \beta$ arbitrary functions.

\subsection{Weak spherical symmetry}
Now, the three generators of spatial rotations SO(3) in a representation
using polar coordinates $x^1= r, x^{2} = \theta, x^{3} = \phi$ are added. Its 
corresponding generators are: \begin{equation}  \xi_{(2)}^{s}= 
\delta_{3}^{s}, \xi_{(3)}= -sin x^3 \delta_{2}^{s} - cos x^{3} ctg
x^{2}~ \delta_{3}^{s}, \xi_{(4)}=cos x^3 \delta_{2}^{s} - sin x^{3} ctg  ~x^{2}
  \delta_{3}^{s}~.\end{equation} Lie-dragging with the time
translation and with $ \xi_{(2)}$ forming the abelian subgroup leads to  
$ \overset{1}{\gamma}_{ab}=g_{ab,0}~,  \overset{2}{\gamma}_{ab}=g_{ab,3},$
and to the weakly Lie invariant metric (i.e., with
${\cal{L}}_{X_1}{\cal{L}}_{X_1}g_{ab} =
0,~{\cal{L}}_{X_{2}}{\cal{L}}_{X_{2}}g_{ab} = 0)$ \begin{equation}  
  g_{ab} = x^0 x^3 c_{ab}(x^1,x^2) + x^0 d_{ab}(x^1,x^2) + x^3
  e_{ab}(x^1,x^2) + f_{ab}(x^1,x^2) \label{metg4}\end{equation} with
four arbitrary bilinear forms $ c_{ab}, d_{ab}, e_{ab},
f_{ab}$.

Lie-dragging with  $\xi_{(3)}$ and  $\xi_{(4)}$
applied to any of these bilinear forms results in the following
equations (using $f_{ab}$ for the presentation):\begin{eqnarray}
  \overset{3}{\gamma}_{ab}= 
  -sin x^3 f_{ab,2} - 2 cos x^3 f_{2(a}\delta_{b)}^{3} + 2 sin x^{3}
  ctg x^{2}f_{3(a}\delta_{b)}^{3} + 2\frac{cos x^3}{sin^2
    x^2}f_{3(a}\delta_{b)}^{2}~,\\ \overset{4}{\gamma}_{ab}= cos x^3
  f_{ab,2} - 2 sin x^3 f_{2(a}\delta_{b)}^{3} - 2 cos x^{3} ctg
  x^{2}f_{3(a}\delta_{b)}^{3} + 2\frac{sin x^3}{sin^2
    x^2}f_{3(a}\delta_{b)}^{2}~.~~\end{eqnarray} The demand
$\overset{2}{\gamma}_{ab}= \overset{3}{\gamma}_{ab}= \overset{4}{\gamma}_{ab}=0 $,
i.e., that {\em spherical symmetry} hold, leads to $c_{ab}=e_{ab}=0$ and to the 
well-known result for $f_{ab}, d_{ab}$: \begin{equation}f_{ab}=
  \alpha(x^1)\delta_{a}^{0}   \delta_{b}^{0} -
  \beta(x^1)\delta_{a}^{1} \delta_{b}^{1}-
  \epsilon(x^1)[\delta_{a}^{2} \delta_{b}^{2} + sin^2 x^{2} 
  \delta_{a}^{3} \delta_{b}^{3}]\label{sssol}\end{equation} with two free
functions $\alpha(x^1), \epsilon(x^1)$.\footnote{One of the functions
  $\alpha(x^1), \beta(x^1)$ is superfluous because, locally, a
  2-dimensional space is conformally flat. $f_{01}=f_{23}=0$ follows
  from the rotation group acting on a 2-dimensional subspace. In addition, here 
  $f_{02}= f_{03}= f_{12}= f_{13}= 0$ has been used.}\\  
  
  If definition 3 for complete sets of weak symmetry is applied up: two further PDE's must 
  then be satisfied. If all generators of the rotation group are taken into account, then the 
  result is   \begin{equation} \gamma_{ab} = x^0 d_{ab}(x^1,x^2) + 
  f_{ab}(x^1,x^2) \label{metg5}\end{equation} with two bilinear forms
$ d_{ab}, f_{ab}$ having the same form: \begin{equation}f_{ab}=  
  \alpha(x^1)\delta_{a}^{0} \delta_{b}^{0} - \beta(x^1)\delta_{a}^{1}
  \delta_{b}^{1}- [x^{2}\epsilon_{1}(x^1)+
  \epsilon_{2}(x^1)][\delta_{a}^{2} \delta_{b}^{2} + sin^2 x^{2}
  \delta_{a}^{3} \delta_{b}^{3}]~\label{nsssol}.\end{equation} For the
proof, we do not reproduce here the lengthy full expressions for \linebreak
${\cal{L}}_{\xi_{(3)}} {\cal{L}}_{\xi_{(3)}} f_{ab} =0 $ and
${\cal{L}}_{\xi_{(4)}} {\cal{L}}_{\xi_{(4)}} f_{ab} =0$, but give only
the equations for the components $f_{22}, f_{33}$:  
\begin{eqnarray}{\cal{L}}_{\xi_{(3)}} {\cal{L}}_{\xi_{(3)}} f_{22} =
  -sin^2 x^2 f_{22,2,2} + 2\frac{cos^{2} x^3}{sin^2 x^2}[-f_{22} +
  \frac{f_{33}}{sin^2 x^2}] \overset{!}{=} 0~,\\{\cal{L}}_{\xi_{(4)}}
  {\cal{L}}_{\xi_{(4)}} f_{22} = -cos^2 x^2 f_{22,2,2} +
  2\frac{sin^{2} x^3}{sin^2 x^2}[-f_{22} + \frac{f_{33}}{sin^2
    x^2}] \overset{!}{=} 0~. \end{eqnarray} The consequences $ f_{22,2,2}=0$ and
$f_{33}=sin^2 x^2 f_{22}$ are obvious. That (\ref{nsssol}) is a
genuine solution is shown by $\gamma_{22}= {\cal{L}}_{\xi_{(3)}} f_{22}=
-sinx^3~ \epsilon_1 (x^1)\neq 0$ and by $\gamma_{33}= {\cal{L}}_{\xi_{(3)}} f_{33}=
-sinx^3 sin^2 x^2~ \epsilon_1 (x^1)\neq 0$ if $\epsilon_1 (x^1)\neq 0.$

The surface $x^1= const, x^0= const$ has Gaussian curvature: \begin{equation} 
K = \frac{1}{2(\epsilon_1 x^2 + \epsilon_2)^2}[-\epsilon_1 ctg x^2 +
2\epsilon_1 x^2  + 2   \epsilon_2 +\frac{(\epsilon_1)^2}{\epsilon_1 x^2 +
    \epsilon_2}].\end{equation} $\epsilon_1, \epsilon_2 $ are 
 now constants. For $\epsilon_1 \rightarrow 0$ we obtain the constant
 curvature of the 2-sphere.\\ 
The time translation and the 3 generators of the rotation group form a
complete set of weak Lie motions; this shows that definition 3 is not
empty. 

However, if it is asked that the rotation group generate a {\em strong} set of
weak symmetries according to definition 2, then the result is very
restrictive. The conditions ${\cal{L}}_{\xi_{(2)}} {\cal{L}}_{\xi_{(3)}} f_{ab}
= 0 = {\cal{L}}_{\xi_{(2)}} {\cal{L}}_{\xi_{(4)}} f_{ab}$ for
equations (\ref{metg4}), (\ref{nsssol}) are leading to the remaining metric
tensor of (\ref{metg5}. If $ {\cal{L}}_{\xi_{(3)}}
{\cal{L}}_{\xi_{(4)}}\gamma_{33} = 0 $ is studied for $f_{ab}$, then $
{\cal{L}}_{\xi_{(3)}} {\cal{L}}_{\xi_{(4)}} f_{ab} \neq 0 $ due to the
only nonvanishing expression ${\cal{L}}_{\xi_{(3)}}
{\cal{L}}_{\xi_{(4)}} f_{33} = sin x^2 cos x^2 \times \epsilon_1(x^1)$ 
for $\epsilon_1(x^1)\neq 0.$ Thus the demand that the rotation group
in 3 dimensions generates a {\em strong} set of weak Lie symmetries according 
to definition 2 enforces $\epsilon_{1}(x^1)= 0$ and reduces to an isometry. 
Nevertheless, the resulting spherically symmetric metric is only
weakly static. \\

\subsection{The group $G_3$ acting as a group of weak Lie motions}
\label{subsection:G3weak}
In taking up the example of a $G_3$ acting on $V_3$ from section
\ref{section:Liedrag} with Lie algebra (\ref{LieG3}), we first apply
definition 3 to a scalar $f(x^1, x^2, x^3)$. If the generators are to
lead to motions, then the only solution is $f= constant$. Definition 3
for a complete set of weak Lie motions leads to:\footnote{The
  calculations are sketched in appendix 2.} \begin{equation} f= a_0
  x^2 x^3 + b_0 x^{1} (x^2-x^1 x^3) + c_0 x^1 x^3 + b_1x^2 + c_1 x^3 + d_1 x^1 +
  d_0~, \label{eqdef3}\end{equation} while definition 2 results  
in: \begin{equation} f=   c_0 (x^1 x^3 + x^2) + c_1 x^3 + d_1 x^1 + 
  d_0~.\label{eqdef4}\end{equation} We note, that the only one of the
9 possible demands so far unused, i.e.,
${\cal{L}}_{\xi_{(3)}}{\cal{L}}_{\xi_{(2)}}g_{ab} = 0 $ reduces
(\ref{eqdef4}) to \begin{equation} f= c_1 x^3 + d_1 x^1 +
  d_0~.\label{eqdef5} \end{equation}  
Applying $G_3$ to the metric, the following weakly Lie-invariant
metric is obtained: \begin{gather*} 
  {\gamma}_{ab}=\\ x^{2} \begin{pmatrix} \overset{(0)}{\alpha}_{11} &
    \overset{(0)}{\alpha}_{12} & P_1 \\ \overset{(0)}{\alpha}_{21} &
    \overset{(0)}{\alpha}_{22} &     P'_1 \\ P_1 & P'_1 &
    P_2 \end{pmatrix}  + x^{3} [~ x^{1}\begin{pmatrix}
    \overset{(0)}{\alpha}_{11} & \overset{(0)}{\alpha}_{12} & P_1 \\
    \overset{(0)}{\alpha}_{21} & \overset{(0)}{\alpha}_{22} &     P'_1
    \\ P_1 & P'_1 & P_2 \end{pmatrix} + 
\begin{pmatrix}\overset{(0)}{\beta}_{11} & \overset{(0)}{\beta}_{12}
& \tilde{P}_1 \\ \overset{(0)}{\beta}_{21} & \overset{(0)}{\beta}_{22}
& \tilde{P'}_1 \\ \tilde{P}_1 & \tilde{P'}_1 &
\tilde{P}_2 \end{pmatrix}]  + \begin{pmatrix} Q_1 & Q_1 & Q_2 \\ Q_1 &
\tilde{Q}_1 & Q_2 \\ 
    Q_2 & Q'_2 & Q_3 \end{pmatrix}, \label{G3weak}\end{gather*} where
$P_i, P_i, Q_i, \tilde{Q_i}, Q_i $ are polynomials in the coordinate $x^1$ of
order $i$, the coefficients of which are not all independent:\\  
$P_1 =\overset{(0)}{\alpha}_{12} x^1 +c_{13}, P_1'=
\overset{(0)}{\alpha}_{22} x^1 + c_{23}, P_2 =
\overset{(0)}{\alpha}_{22} (x^{1})^2 + 2c_{23}x^1 + c_{33},\\ Q_1=
\overset{(0)}{l}_{11}x^1+\overset{(0)}{m}_{11}, Q_1=
\overset{(0)}{l}_{12}x^1+\overset{(0)}{m}_{12}, \tilde{Q}_1=
{l}_{22}x^1+\overset{(0)}{m}_{22}, \\Q_2=
\overset{(0)}{l}_{12}(x^1)^2+\overset{(0)}{m}_{13}x^1 +
\overset{(0)}{k}_{13},~Q_2=
\overset{(0)}{l}_{22}(x^1)^2+\overset{(0)}{m}_{23}x^1 +
\overset{(0)}{k}_{23},\\ Q_3=
\overset{(0)}{l}_{22}(x^1)^3+\overset{(0)}{m}_{23}(x^1)^2 +
\overset{(0)}{k}_{33}x^1 + m_{33} $ and $\overset{(0)}{\alpha}_{ab},
(a, b = 1, 2),~ c_{ij},  \overset{(0)}{l}_{ij},~
\overset{(0)}{m}_{ij}$ and $ \overset{(0)}{k}_{ij}$ constants. In the 
polynomials $\tilde{P}_1,\tilde{P'}_1,\tilde{P}_2,$ the constants
$\alpha_{ab}, c_{ab}$ are exchanged by the set of independent constants
$\beta_{ab}, d_{ab}$. Thus definition 2 is not empty. Two independent
matrices of the type that occured for the group acting as an 
isometry group and a third, new matrix occur now.\\ 
Definition 3 leads to a different complete set of weak Lie motions for which 
the metric takes the form:
$g_{ab} = d_{ab}(x^1) x^2 + e_{ab}(x^1) x^3 +  \epsilon_{ab}(x^1)$ with
$d, e, \epsilon$ expressed by matrices of the form:
\begin{gather*} \begin{pmatrix} P_{11} & P_{12} & Q_1\\ P_{12} & P_{22} & Q_2\\
 Q_1 & Q_2 & M  \end{pmatrix}, \label{G3weak2}\end{gather*} where
 $P_{ik}$ are polynomials of 1st degree, $Q_i$ of 2nd degree, and $M$ a
 polynomial of 3rd degree.

\section{A new algebra structure}
\label{section:newalgeb} 
For Lie-dragging, up to now we have mostly taken vector fields forming
Lie algebras corresponding to Lie groups of point transformations. In
the following, after an introductory section, we consider more general types 
both of groups and algebras in sections \ref{section:algebext} and
\ref{section:extmotion}. 

\subsection{Lie-dragging for vector fields not forming Lie 
algebras}
\label{subsection: non-Lie}
Already in (\ref{freefunc}) of section \ref{subsubsection:weaksym},
vector fields containing free functions were considered. We now continue
with vector fields $X_1=\xi^{r}\frac{\partial}{\partial x^{r}}; ~X_2=
\eta^{s}\frac{\partial}{\partial x^{s}}$ with $ \xi^{r}=
f(x^0)\delta_{1}^{r}, \eta^{s}= h(x^{1})\delta_{0}^{r}$ such
that \begin{equation}[X_1, X_2] =  f(x^0) H(x^{1}) X_2 - h(x^{1})
  F(x^0) X_1~.\label{extLiealg}\end{equation} Here, $ F(x^0)=
\frac{d(ln f(x^{0}))}{dx^{0}},  H(x^{1})= \frac{d(ln
  h(x^{1}))}{dx^{1}}$. The finite transformations belonging to  $X_1$
and $X_2$, respectively, are generalized time- and space-translations 
\begin{equation}x^0 \rightarrow x^{0'} = x^0 +
  h(x^1) ; ~x^1 \rightarrow x^{1'} = x^1 +
  f(x^0)~ \label{trafo}\end{equation} leaving invariant the time interval 
$|x^{0}_{(i)} -x^{0}_{(j)}|$ and the space interval $|x^{1}_{(i)} - x^{1}_{(j)}|$
between two events $(x^{0}_{(i)},x^{1}_{(i)})$ and $(x^{0}_{(j)},x^{1}_{(j)})$. 
Each of the transformations
\begin{equation} x^0 \rightarrow x^{0'} = x^0 + h(x^1) ; ~x^1 \rightarrow x^{1'} =
x^1 + a ,\label{trafospez1}\end{equation} and \begin{equation} x^1
\rightarrow x^{1'} = x^1 + f(x^0) ; ~x^0 \rightarrow x^{0'} =
x^0 + b\label{trafospez2}\end{equation} forms a group: $
x^{0''} = x^{0'} + k(x^{1'}) = x^0 + h(x^1) + k(x^{1}+a),~ x^{1''} =
x^{1}+ A+ a$, and $ x^{1''} = x^{1'} + g(x^{0'}) =  x^{1} + f(x^0) +
g(x^0+b),~ x^{0''} = x^{0} + B +b$. However, these groups are {\em
  not} Lie groups: in part, the Lie-group parameters have been
replaced by arbitrary functions. In this case, the algebra
(\ref{extLiealg}) reduces to either \begin{equation}[X_1, X_2] =  H(x^1)
  X_2~. \label{specextLiealg1}\end{equation} or
to \begin{equation}[X_1, X_2] =  - F(x^0)
  X_1~. \label{specextLiealg2}\end{equation} Likewise,
(\ref{extLiealg}), (\ref{specextLiealg1}), and (\ref{specextLiealg2})
are not Lie algebras. 

Both transformations (\ref{trafo}) applied together: $ x^{0''} =
x^{0'} + k(x^{1'}) =  x^0 + h(x^1) + k(x^{1} + f(x^{0})), ~  x^{1''} =
x^{1'} + g(x^{0'}) =  x^1 + f(x^0) + g(x^{0} + h(x^{1}))$ do not even form a group. 
 
 The class of functions involved may be  narrowed considerably by the
 demand that the function $f$ of a special type be kept fixed, e.g.,
 be a polynomial of degree $p$, or $f(x^0) = a sin x^0 +  b cos x^0 $. In
 these cases, just one function with constant coefficients occurs in
 the group; the group transformations change only the
 coefficients. (\ref{trafospez2}) is a  subgroup of the so-called 
{\em Mach-Poincar\'e group} $G_{4}(3)$ \cite{treder1972},
(\cite{goenner1981}, pp. 85-101):\begin{equation} x^{a'} = 
   A_{~r}^{a} x^{r} + f^{a}(x^{0}),~ x^{0'}= x^{0} + b,~~ A_{~r}^{a}A_{~b}^{r}=
   \delta_{b}^{a}.\end{equation} This group plays a role in Galilean
 relative mechanics. 

A generalization is the group $G_{1}(6)$ of transformations leaving
invariant the observables describing a {\em    rigid body}; 6 free
functions of $x^{0}$ and one Lie-group  parameter do appear: \begin{equation}
  x^{i'} = A_{~j}^{i}(x^{0}) x^{j} +    f^{i}(x^{0}),~ x^{0'}= x^{0} + b,~~
   A_{~j}^{i}(x^{0})A_{~k}^{j}(x^{0})=\delta_{k}^{i},~ (i,j =1,2,3).\end{equation}
 The corresponding  seven algebra generators are:  \begin{eqnarray} T =
   \frac{\partial}{\partial x^{0}},~ X_{i}=
   f_{i}(x^{0})\frac{\partial}{\partial x^{i}}~~(i~ not~
   summed),\nonumber ~~~~~~\\ Y_1 =
   \omega^{2}_{3}(x^{3}\frac{\partial}{\partial x^{2}}-
   x^{2}\frac{\partial}{\partial x^{3}} ), Y_2 =
   \omega^{1}_{3}(x^{3}\frac{\partial}{\partial x^{1}}-
   x^{1}\frac{\partial}{\partial x^{3}}), Y_3 =
   \omega^{1}_{2}(x^{2}\frac{\partial}{\partial x^{1}}-
   x^{1}\frac{\partial}{\partial x^{2}}) \nonumber\\ \end{eqnarray} 
   with $ \omega^{i}_{j}= \omega^{i}_{j}(x^{0})$. The corresponding algebra is
 given by:\begin{eqnarray} [T, T] = 0,~[ T, X_{i}]= F_{i}(x^0) X_{i},~
   F_{i}= \frac{d}{dx^{0}}ln(f_{i} (x^{0})),~[X_{i}, X_{j}]= 0 ,~(i,j
   = 1,2,3)~~~~\nonumber\\~ [T, Y_1]= \omega^{2}_{3}(x^{0})Y_1,~[T,
   Y_2]= \omega^{3}_{1}(x^{0})Y_2, ~[T, Y_3]= 
   \omega^{1}_{2}(x^{0})Y_3,~
   \omega^{i}_{j}=\frac{d}{dx^{0}}ln(\omega^{i}_{j}(x^{0})),\nonumber\\~ [Y_1, Y_2]=
   -\frac{\omega^{1}_{3}\omega^{2}_{3}}{\omega^{1}_{2}}~Y_3,~ [Y_2, Y_3]=
   -\frac{\omega^{1}_{2}\omega^{1}_{3}}{\omega^{2}_{3}}~Y_1,~ [Y_1, Y_3]=
   -\frac{\omega^{1}_{2}\omega^{3}_{2}}{\omega^{1}_{3}}~Y_2,~~~~~\nonumber
   \\~ [X_1, Y_1] = 0,~ [X_1, Y_2]=
   \frac{f_1(x^{0})}{f_3(x^{0})}\omega^{1}_{3}~X_3,~  
   [X_1, Y_3] = - \frac{f_1(x^{0})}{f_2(x^{0})}\omega^{1}_{2}~X_2,~~~~~\nonumber
   \\~[X_2, Y_1]= \frac{f_2(x^{0})}{f_3(x^{0})}\omega^{2}_{3}~X_3,~
 [X_2, Y_2] = 0,~ [X_2, Y_3] =
 \frac{f_2(x^{0})}{f_1(x^{0})}\omega^{1}_{2}~X_1, ~~~~~~~\nonumber \\~
 [X_3, Y_1]= \frac{f_3(x^{0})}{f_2(x^{0})}\omega^{2}_{3}~X_2,~ [X_3,
 Y_2] = - \frac{f_3(x^{0})}{f_1(x^{0})}\omega^{1}_{3}~X_1,~  [X_3,
 Y_3] = 0 .~~~~~~ \label{rigbody}
 \end{eqnarray} There exist further groups of this non-Lie type
 occuring in classical mechanics like Weyl's kinematical group
 $G_{3}(6)$ and the covariance group of the Hamilton-Jacobi equation
 $G_{7}(3)$ or, as a subgroup in non-relativistic quantum mechanics,
 the covariance  group of the Schr\"odinger equation $G_{12}(0)$,
cf. \cite{goenner1981}. The structure functions of all these groups depend on
 a single coordinate, the time.

\section{Extended Lie Algebras}
\label{section:algebext} In the following, we will deal with a
subbundle of the tangent bundle of n-dimensional Euclidean or Lorentz
space. We will permit that the structure constants in the defining
relations for a Lie algebra become dependent on the components
$\xi^{a}_{~i}$ of the vector fields $X_{i}(x)$: they will become {\em
structure functions}.\\ 

{\em Definition 4}:\\ 
The algebra \begin{equation} [X_i, X_j ]= c_{ij}^{~k}( x^1, x^2, ...,
  r) X_k \label{Killing}\end{equation} with structure functions 
$c_{ij}^{~k}(x^1, x^2, ...,x^{r})$ is called an {\em extended Lie algebra}.\\

 The Lie algebra elements form an ``involutive distribution''. This is
 ``a smooth distribution $V$ on a smooth manifold $M$, i.e., a smooth
 vector subbundle of the tangent bundle'' $TM$. The Lie brackets
 constitute the composition law; the injection $V  \hookrightarrow TM$
 functions as the anchor map (cf. \cite{Marle2008}, p. 13). This is a
 simple example for a {\em tangent Lie algebroid} (cf. also
 (\cite{Mack2005}, p. 100  and example 2.7, p. 105)).\footnote{Closely
related, but different structures are {\em family of Lie algebras}
   \cite{DouLaz66}, \cite{Copp77} and {\em variable Lie algebras}
   (\cite{LepLud94}, p. 115).} Nevertheless, the involutive
 distribution used here can also be considered a subset of the
 infinite-dimensional ``Lie-algebra''  ${\cal B}(M)$ of footnote 5.
 
 After completion of the paper, I learned of some of the historical 
 background  of (\ref{Killing}): It already has occured as the condition for 
 closure of a complete set of linear, homogeneous operators belonging
 to a complete system of 1st order PDE's in Jacobi's famous paper of 1862 
(\cite{Jacobi1862}, \S 26, p. 40).\footnote{In Jacobi's paper,
(\ref{Killing}) is used in phase space such that the structure
functions depend on both coordinates and momenta: $c_{ij}^{~k}(x^1,
x^2, ...,x^{r}, p_1, p_2, ..p_r)$. It is in Clebsch's paper of 1866
(\cite{Clebsch1866}, \S 1) in connection with his definition of a complete
system of linear PDE's that the r.h.s. of (\ref{Killing}) depends only on the 
coordinates. Cf. also equation (3.1) in \cite{Hawk1989}, p. 311.} \\  

(\ref{Liealg2}) must then be replaced
by \begin{equation}{\cal{L}}_{X_{i}} {\cal{L}}_{X_{j}} X_{k} =
  (c_{jk}^{~~l} c_{il}^{~~m} + X_{i} c_{jk}^{~~m})
  X_{m}~,\label{Liealg3}\end{equation} and (\ref{Jaco2})
by \begin{equation} c_{jk}^{~~l} c_{il}^{~~m} +c_{ij}^{~~l}
  c_{kl}^{~~m}+c_{ki}^{~~l} c_{jl}^{~~m} +  X_{i} c_{jk}^{~~m} + X_{k}
  c_{ij}^{~~m} +  X_{j} c_{ki}^{~~m}=0. \label{Jaco3}\end{equation}
An {\em extended Cartan-Killing} form can be defined acting as a symmetric 
metric on the sections of the subtangent bundle. An asymmetric form 
could be defined as well. \\

Definition 5 (Generalized Cartan-Killing form):\\
The generalized Cartan-Killing bilinear form $\tau$ is defined
by: \begin{equation} \tau_{ij}:= \sigma_{ij} + 2 X_{(i} c_{j)m}^{~~~m} =
  c_{il}^{~~m}c_{jm}^{~~~l} + 2 X_{(i} c_{j)m}^{~~~m}~.\label{CaKiext}\end{equation} 

The generalized Cartan-Killing form now depends on the base points of the fibres 
in the tangent bundle. They may be interpreted as a metric.\\ 

To use the example of the group $G_{1}(6)$ given in section
\ref{subsection: non-Lie}: The structure functions for the
corresponding extended algebra (\ref{rigbody}) of rigid body
transformations are shown in appendix 4. From them, calculation of the 
extended Cartan-Killing form leads to a Lorentz metric with signature
(1,3) of rank 4 within a degenerated 7-dimensional bilinear form:   
 \begin{gather} \tau_{ij}= \begin{pmatrix} \tau_{00} & 0 &
     0 & 0 & 0 & 0 & 0 \\
0 & 0 & 0 & 0 & 0 & 0 & 0 \\ 0 & 0 & 0 & 0 & 0 & 0 &
0\\ 0 & 0 & 0 & 0 & 0 & 0 & 0 \\ 0 & 0 & 0 &
0 & \tau_{44} & 0 & 0 \\ 0 & 0 & 0 & 0 & 0 & \tau_{55}
& 0 \\ 0 & 0 & 0 & 0 & 0 & 0 &\tau_{66} \\ \end{pmatrix} \end{gather}  
where $\tau_{00}= \Sigma_{i=1}^{3}\frac{\ddot{f}_{i}}{f_{i}}+
\frac{\ddot{\omega}_{~3}^{2}}{\omega_{~3}^{2}} +
\frac{\ddot{\omega}_{~1}^{3}}{\omega_{~1}^{3}}+
\frac{\ddot{\omega}_{~2}^{1}}{\omega_{~2}^{1}}$, and $\tau_{44}=
-4(\omega_{~3}^{2})^2,~ \tau_{55}= -4(\omega_{~3}^{1})^2,~ \tau_{66}=
-4(\omega_{~2}^{1})^2.$  By projection into the 4-dimensional space
with coordinates $0, 4, 5, 6$ and signature (1,3), we surprisingly
arrive at the general class of one-dimensional gravitational fields
\cite{bible}. For special values for the $f_{i},$ and $ \omega_i^k$, the
Kasner metric (\ref{Kasner}) can be derived by this approach. All the
pre-relativistic groups mentioned at the end of the previous section
lead to Cartan-Killing forms depending on just one coordinate, the time.\\
  
The following definition introduces a new class of extended motions and a new class 
of weak extended motions, the infinitesimal generators of which form an extended 
Lie algebra.\\

{\em Definition 6} (extended motions):\\ 
Let $x \rightarrow x+\xi,~ y \rightarrow y + \eta$ be infinitesimal
transformations forming a continuous group the corresponding
algebra of which is an extended Lie algebra according to definition
4. Then, the vector fields $X=\xi^{c} \frac{\partial}{\partial x^{c}},~Y = \eta^{c}  \frac{\partial}{\partial x^{c}}$ with ${\cal L}_{X} g_{ab}=0,~{\cal L}_{Y} g_{ab}=0$ are called {\em extended motions}.\\

An analogous formulation is:\\
{\em Definition 7} (extended weak motions):\\ 
Let $x \rightarrow x+\xi,~ y \rightarrow y + \eta$ be infinitesimal
transformations forming a continuous group the corresponding
algebra of which is an extended Lie algebra according to definition
4. Then, the vector fields $X=\xi^{c} \frac{\partial}{\partial x^{c}},~Y = \eta^{c}  \frac{\partial}{\partial x^{c}}$ with ${\cal L}_{X} {\cal L}_{X}g_{ab}=0,~{\cal L}_{Y}{\cal L}_{Y} g_{ab}=0$ are called {\em extended weak motions}.\\

\section{Extended motions and extended weak (Lie) motions}
\label{section:extmotion}
In section (\ref{subsection: non-Lie}), we have given examples
of non-Lie groups leading to extended Lie algebras. How will the
corresponding extended motions and extended weak (Lie) motions differ? 
These concepts are exemplified here with the  most simple non-Lie group
(\ref{trafospez2}). The tangent vectors $X,~Y$ with the algebra
(\ref{specextLiealg2}) form an extended motion (${\cal L}_{X}
g_{ab}=0,~{\cal L}_{Y} g_{ab}=0$) for all metrics of maximal rank 3:
\begin{gather}
  g_{ab}= \begin{pmatrix} 
    \alpha_{00}(x^2, x^3) & 0  & \alpha_{02}(x^2, x^3) & \alpha_{03}(x^2, x^3) \\
   0 & 0 & 0 & 0 \\ \alpha_{02}(x^2, x^3) & 0 &  \alpha_{22}(x^2, x^3) & \alpha_{23}(x^2, x^3) \\ 
      \alpha_{03}(x^2, x^3) & 0 & \alpha_{23}(x^2, x^3) & \alpha_{33}(x^2, x^3) \end{pmatrix}, \label{G3weak2}\end{gather} with arbitrary functions $\alpha_{ab}$ due to arbitrariness of $f(x^0)$. This is to be compared with the motions derived from  $X=\frac{\partial}{\partial x^{1}}, Y=\frac{\partial}{\partial x^{0}} $ forming an abelian Lie algebra and leading to \begin{equation} g_{ab} = \alpha_{ab}( x^2, x^3)~.\end{equation}

The corresponding extended weak (Lie) motions  (${\cal L}_{X} {\cal L}_{X}g_{ab}=0,~{\cal L}_{Y}{\cal L}_{Y} g_{ab}=0$) are given by:
\begin{gather}
  g_{ab}= \begin{pmatrix} 
     x^1\alpha_{00} +\beta_{00}  & \beta_{01} & x^1\alpha_{02} + \beta_{02} & x^1\alpha_{03} + \beta_{03}\\ \beta_{01} & 0 &  \beta_{12} &  \beta_{13} \\ x^1\alpha_{02} + \beta_{02} & \beta_{12} &   x^1\alpha_{22} + \beta_{22} & x^1\alpha_{23} + \beta_{23} \\ x^1\alpha_{03} + \beta_{03} &  \beta_{13}  & x^1\alpha_{23}+ \beta_{23} & x^1\alpha_{33}+ \beta_{33} \end{pmatrix}, \label{G3weak3}\end{gather} where $\alpha_{ab}=\alpha_{ab}( x^2, x^3);~\beta_{ab}=\beta_{ab}( x^2, x^3),\alpha_{0a} = 0, \beta_{11}= 0$. Comparison with the weak (Lie) motions generated by the translations given above shows
the class of metrics: \begin{equation} g_{ab} = x^1\alpha_{ab}( x^2, x^3) + \beta_{ab}( x^2, x^3)~.\end{equation}    

\section{Two-dimensional extended Lie algebras}
 \label{section:classextalg}
In section \ref{subsection: non-Lie} we have given the example
(\ref{extLiealg}) showing that (\ref{Killing}) is not empty.
As for Lie algebras, the question about a classification of
extended Lie algebras in n-dimensional space arises. This being a
topic of its own, we start here by considering the case $n=2$ only,
without proving completeness of the result.
 
We begin with:\footnote{The coordinates $x^0, x^1$ of section \ref{subsection: non-Lie} are
  replaced by  $x^1, x^2$.} \begin{equation} [X_{1}, X_{2}]= c_{12}^{~~1} X_{1} +
  c_{12}^{~~2} X_{2} \label{starteq}\end{equation} with
$X_{1}=\xi^{1}\frac{\partial}{\partial x_1} + 
\xi^{2}\frac{\partial}{\partial x_2}, X_{2} =
\eta^{1}\frac{\partial}{\partial x_1} +
\eta^{2}\frac{\partial}{\partial x_2}.$ This is a system of two
equations for the 6 unknowns $\xi^i, \eta^j$ and $c_{12}^{~~i}, i= 1,
2$:\begin{equation} [X_{1}, X_{2}]= [\xi^1 \eta^1_{~,1}+ \xi^2
  \eta^1_{~,2} - \eta^1 \xi^1_{~,1} -\eta^2 \xi^1_{~,2}]
  \frac{\partial}{\partial x_1} + [\xi^1 \eta^2_{~,1}+ \xi^2
  \eta^2_{~,2} - \eta^1 \xi^2_{~,1} -\eta^2 \xi^2_{~,2}]
  \frac{\partial}{\partial x_2}.~ \label{eq2ext}\end{equation} We
distinguish two cases according to whether the vector fields are
unaligned or aligned. 
In the {\em first case}, for $\xi^{1}\neq 0,~\eta^{2}\neq 0$  :\begin{equation}
  [X_{1}, X_{2}]= [(\xi^1)^2 \frac{\partial}{\partial x_1}
  (\frac{\eta^1}{\xi^1}) + \xi^2 \eta^1_{~,2}- \eta^2 \xi^1_{~,2}]
  \frac{\partial}{\partial x_1} + [\xi^1 \eta^2_{~,1}- \eta^1
  \xi^2_{~,1} - (\eta^2)^2 \frac{\partial}{\partial x_2} (\frac{\xi^2}{\eta^2})
  \frac{\partial}{\partial x_2}]~.\label{eq2ext1} \end{equation}  
Here, the simplification $\xi^{2} = \eta^{1}= 0$ does not restrict
generality. In the solution, two free functions $\xi^{1}, \eta^2$
remain; they are contained in the expressions for the structure
functions:\begin{equation}c_{12}^{~~1}=-\frac{\eta^2}{\xi^1} 
  \xi^{1}_{~,2}~,~c_{12}^{~~2}= \frac{\xi^1}{\eta^2} \eta^{2}_{~,1}~. \end{equation}
The further simplification $\xi^1=\eta^2$ leads to:\begin{equation}
  [X_{1}, X_{2}]=   -\xi^{1}_{~,2} X_{1} + \xi^{1}_{~,1}
  X_{2} \label{example1}\end{equation} with arbitrary $\xi^{1}=\xi^{1}
(x^1, x^2).$ Calculation of the extended Cartan-Killing form
(\ref{CaKiext}) results in: \begin{gather*}\tau_{ik}=\begin{pmatrix}
    ~([\xi^{1}_{~,1})^2] + \xi^{1}\xi^{1}_{~,1,1} &
    \xi^{1}_{~,1}\xi^{1}_{~,2}+\xi^{1}\xi^{1}_{~,1,2} ~\\ ~
    \xi^{1}_{~,1}\xi^{1}_{~,2}+\xi^{1}\xi^{1}_{~,1,2} & (\xi^{1}_{~,2})^2 +
    \xi^{1}\xi^{1}_{~,2,2} \end{pmatrix} ~,\end{gather*} or
simply \begin{equation} \tau_{ij}= \frac{1}{2} [
  (\xi^{1})^{2}]_{,ij}~.\end{equation} In general $det(\tau_{ik})\neq
0$.\\

In order to find (\ref{specextLiealg2}) in this formalism, we must
start from (\ref{trafospez2}) and set $\xi^1= f(x^0), \eta^0= 1$ such
that $c_{12}^{~~1}= -F(x^0),~ c_{12}^{~~2} = 0$. As the only dependence
is on $x^0$, the Cartan-Killing form degenerates (does not have full
rank). This also happens for the algebra (\ref{rigbody}).\\

For 2-dimensional Lorentz space, one of the generators can be
lightlike. We use only the simplification $\xi^{2} = 0$ and $\eta^{1}=
\pm \eta^{2}$ such that in this case the relation: \begin{equation}
  [X_{1}, X_{2}]= [\frac{\pm 1}{(\xi^1)^2} \frac{\partial}{\partial
    x_1} (\frac{\eta^1}{\xi^1}) - \eta^2 \xi^1_{~,2}]
  \frac{\partial}{\partial x_1}  + \xi^1 \eta^2_{~,1}
  \frac{\partial}{\partial x_2}\end{equation} follows. Again, we can
set  $\xi^1=\eta^2$ and come back to (\ref{example1}). 

The two different Lie algebras allowed in 2-dimensional space can be obtained
from (\ref{example1}) by special choice of $\xi^1$. By redefinition of
the algebra elements in the sense of \begin{equation} X_1 \rightarrow
  Y_1 = f(x^1, x^2) X_1 +  g(x^1, x^2) X_2~,~ X_2 \rightarrow Y_2 =
  m(x^1, x^2) X_1 +  p(x^1, x^2) X_2~\end{equation} with arbitrary
functions $f, g, m, p$,  from (\ref{starteq}) it may be possible to come back to
the canonical form for the non-abelian Lie algebra. However, this is an open 
question.\footnote{It depends on whether solutions of certain nonlinear 1st order 
PDEs exist.}\\ 

In the {\em second case} of aligned tangent vectors we can set
$\xi^2=0=\eta^2$. From (\ref{eq2ext}) we retain as the only structure
function: \begin{equation} c_{12}^{~~1} = \xi^{1} \eta^{1}_{~,1}-
  \eta^{1} \xi^{1}_{~,1}~.\end{equation} The Cartan-Killing form then
is: \begin{gather}\tau_{ik}=\begin{pmatrix} 0 & - \xi_{1} (\xi^{1} \eta^{1}_{~,1}-
  \eta^{1} \xi^{1}_{~,1})\\  - \xi_{1} (\xi^{1} \eta^{1}_{~,1}-
  \eta^{1} \xi^{1}_{~,1})~ &~ (\xi^{1} \eta^{1}_{~,1}-
  \eta^{1} \xi^{1}_{~,1})^{2}-\eta_{1}( \xi^{1} \eta^{1}_{~,1,1}-
  \eta^{1} \xi^{1}_{~,1,1})~ \end{pmatrix} ~.\end{gather}\\

In general, there is a wealth of possibilities available for setting up
extended Lie algebras. A particular choice for the structure functions
would be\footnote{In (\ref{Killing2}), the summation convention is
  used for unbracketed indices.} \begin{equation}c_{(i)
    (j)}^{~~~~(k)}:=   \xi^{r}_{(i)} \xi_{(j)}^{s} g_{rs}
  [\delta^{(k)}_{(i)} -
  \delta^{(k)}_{(j)}] \label{Killing2}~.\end{equation} In Euclidean
space $ g_{rs}= \delta_{rs}$, in Lorentz space $ g_{rs}= \eta_{rs}$
are most simple choices. It is not difficult to calculate the extended
Cartan-Killing form which depends only on the inner products $~
\xi^{r}_{(i)}  \xi_{(j)}^{s} g_{rs}, (i), (j) = 1, 2, , ..., m;~ r,s =
1, 2, ..., n ~(0, 1, 2, ..., n-1)$.\\  

\section{Discussion and conclusion}
When Lie-dragging is seen as a mapping in the space of metrics, it may be asked
whether it could provide a method for generating solutions of
Einsteins equations from known solutions. It is easily shown that the
Schwarzschild vacuum solution, the Robertson-Walker metric with flat
3-spaces, and the Kasner metric cannot be obtained by Lie-dragging of
{\em Minkowski} space. On the other hand, the metric (\ref{metg4})
which is weakly Lie-invariant with respect to the group  (T, SO(3))
trivially contains cosmological solutions of Einsteins equation. If
the metric $x^0 d_{ab}(x^1, x^2) $ with spherically symmetry and with
flat space sections is chosen, by a transformation of the time
coordinate we arrive at the line element \begin{equation} ds^2 =
  (d\tau)^2 - 2/3 \tau^{3/2}[(dr)^2 
  + r^2 (d\theta)^2 + r^2 sin^2\theta (d\phi)^2]~.\end{equation} It
describes a cosmic substrate with the equation of state
$p=-\frac{1}{9}\mu$, where $\mu$ describes pressure and $\mu$ the
energy density of the material. This equation of state for
$w=-\frac{1}{9}$ is non-phantom because of $-1< w$ but does not
accelerate the expansion of the universe which occurs for $-1<
w<-\frac{1}{3}$.

 It remains to be seen  whether the anisotropic line
element \begin{equation} ds^2 = (d\tau)^2 - c_0 \tau^{2/3}[c_1 \theta 
  r + c_2] [(dr)^2 + r^2 (d\theta)^2 + r^2 sin^2\theta
  (d\phi)^2]\end{equation} can satisfy Einstein's equations with a
reasonable matter distribution. In view of the fact that Lie-dragging
does not preserve the rank of the metric, its efficiency for
generating interpretable gravitational fields is reduced considerably.  

Surprisingly, by studying the rigid body transformations $G_{1}(6)$ as a group 
of extended motions, we arrived at the complete class of one-dimensional
gravitational fields including the Kasner metric. More generally, a
close relation to {\em finite} tranformation groups in classical,
non-relativistic mechanics containing arbitrary functions has been established. 
It is still to be cleared up whether a connection to gauge theory in physics exists. 

A classification of solutions of Einstein's equations with regard to weak (Lie) 
symmetries could be made. Although this might be a further help for deciding  
whether two solutions are transformable into each other or not, the calculational 
effort looks extensive. 

Weak Lie-invariance as a weakened concept of ``symmetry'' has been
introduced and its consequences presented through a number of examples. It
also has led to the introduction of a new type of algebra (``extended Lie
algebra'') which is an example for a tangent Lie algebroid. In each {\em fibre} 
of a subbundle of the tangent bundle, the ``extended Lie algebra'' reduces to a 
Lie algebra. By help of an extended Cartan-Killing form, Riemann or Lorentz metrics 
have been constructed on such an algebroid.  A particular example is provided by the 
non-Lie groups of classical mechanics mentioned above. The ensuing possible
geometries could be studied and classified in the spirit of Felix Klein. A classification of non-Lie groups leading to extended Lie algebras and of the extended Lie algebras could also be of interest. A further study of the concept of extended Lie algebras is needed and might be of some relevance.

Whether there are noteworthy applications in geometry and
physics beyond those established here for classical mechanics and the
Schr\"odinger equation will have to be found out. 

\section{Acknowledgments}   
My sincere thanks go to A. Papadopoulos for inviting me to the
Strasbourg conference at IRMA. Remarks by P. Cartier and Y. Kosmann-Schwarzbach, 
participants at the conference, were quite helpful. I am also grateful for advice on
mathematical concepts like algebroids by H. Sepp\"anen and Ch. Zhu of
the Mathematical Institute of the University of G\"ottingen as well as
to my colleague F. M\"uller-Hoissen, Max Planck Institute for Dynamics and
Self-Organization, G\"ottingen, for a clarifying conceptual
discussion. As a historian of mathematics, E. Scholz, University of
Wuppertal, did guide me to the relevant historical literature.

\section{Appendix}
\subsection{Appendix 1 (Integrability conditions)}    
That an {\em arbitrary} symmetrical tensor field of fixed rank {\em
cannot} be reached by the operation of Lie-dragging may be
seen already from (\ref{isomet}), (\ref{examp1}), or from the
following equations obtained from (\ref{isomet2}): \begin{equation}
  \overset{g}{\nabla}_{b} 
  \overset{g}{\nabla}_{c}\xi_{a} + \xi_{d}R^{d}_{~bca}(g)= 1/2~
  [\overset{g}{\nabla}_{b} \gamma_{ac} + \overset{g}{\nabla}_{c}
  \gamma_{ba} + \overset{g}{\nabla}_{a} \gamma_{cb}]~.\label{cond1}\end{equation}
Here, $R^{d}_{~bca}(g)$ is the curvature tensor of the metric
$g_{ab}$. For $g_{ab}, \gamma_{ab}$ fixed, the $n^3$ equations (\ref{cond1}) would 
be an integrability condition for the $n$ components of the vector
field $\xi$. Eq. (\ref{cond1}) generalizes part of the integrability
condition for (\ref{isomet}) given in \cite{Yano57}, eq. (6.2), p. 56. As an
example, we take Minkowski space $g_{ab} = \eta_{ab}$, a space of maximal 
symmetry. From (\ref{cond1}) with $\gamma_{ab}=2\xi_{(a,b)}$
follows: \begin{equation} \partial_b \partial_c \xi_{a} = \partial_b
  \xi_{(a,c)} + \partial_c \xi_{(b,a)} + \partial_a
  \xi_{(c,b)}~, \label{cond1A}\end{equation} the general solution of
which, apart from the generators of the Poincar\'e group, is 
$\xi_{a}= c_0 \eta_{ar}x^{r} + \partial_{r}F^{[rs]} \eta_{as}$
(cf. appendix 3). Thus, a homothetic motion appears as well.

If we look at this equation as a condition for $\gamma_{ab}$ when the
vector field $\xi$ and the metric $g_{ab}$ are given, the equation
then says that there are linear relations between the first derivatives of
$\gamma_{ab}$. Further differentiation of
(\ref{cond1}) leads to:\begin{eqnarray}\overset{g}{\nabla}_{d}
  \overset{g}{\nabla}_{b} \overset{g}{\nabla}_{c}~\xi_{a} +
  \overset{g}{\nabla}_{a} \overset{g}{\nabla}_{b}
  \overset{g}{\nabla}_{d}~\xi_{c} + \overset{g}{\nabla}_{c}
  \overset{g}{\nabla}_{b} \overset{g}{\nabla}_{a}~\xi_{d}= 1/2~ 
  [\overset{g}{\nabla}_{d}\overset{g}{\nabla}_{b} \gamma_{ac} +
  \overset{g}{\nabla}_{a} \overset{g}{\nabla}_{b} \gamma_{cd} +
  \overset{g}{\nabla}_{c} \overset{g}{\nabla}_{b} \gamma_{da}]
  \nonumber \\  +  \gamma_{as}R^{s}_{~bcd}(g) +
  \gamma_{cs}R^{s}_{~bda}(g) + \gamma_{ds}R^{s}_{~bac}(g) = 0
  ~.~~\label{cond2}\end{eqnarray} A counting of derivatives and
equations leads to the number of restrictions for the obtainable
$\gamma_{ab}$ showing up explicitely as relations among the
derivatives of $\gamma_{ab}$. \\

\subsection{Appendix 2}
We find \begin{eqnarray}{\cal{L}}_{X_1}{\cal{L}}_{X_1}= f_{,2,2}= 0,~~
{\cal{L}}_{X_2}{\cal{L}}_{X_2}= f_{,3,3}= 0,\\
{\cal{L}}_{X_3}{\cal{L}}_{X_3}= f_{,1,1} - 2 x^3 f_{,2,1} + (x^3)^2 f_{,2,2} = 0,
~{\cal{L}}_{X_1}{\cal{L}}_{X_2} = {\cal{L}}_{X_2}{\cal{L}}_{X_1} = f_{,2,3} = 0,\\
{\cal{L}}_{X_1}{\cal{L}}_{X_3}= -f_{,1,2} + x^3 f_{,2,2} =
0,~~{\cal{L}}_{X_3}{\cal{L}}_{X_1}= -f_{,1,2} + x^3 f_{,2,2} = 0,~\\
{\cal{L}}_{X_2}{\cal{L}}_{X_3}= -f_{,1,3} + f_{,2} +x^3
f_{,2,3}=0,~~{\cal{L}}_{X_3}{\cal{L}}_{X_2}= -f_{,3,1} + x^3 f_{,3,2}
=0~, \end{eqnarray} from which the results (\ref{eqdef3})-(\ref{eqdef5}) follow.

\subsection{Appendix 3}
From (\ref{cond1A}), by contraction with $\eta^{bc}$ the equation
$\partial_{a} (\partial^{c} \xi_{c}) = 0$ follows, whence
$\partial^{c} \xi_{c}= c_0 = const$. Contraction with $\eta^{ac}$ then
leads to $(\nabla)^{2} \xi_{c}= 0$. The most general Ansatz for
solving  $\partial^{c} \xi_{c}= c_0 $ is  $\xi^{c}= \frac{c_0}{4}
x^{c} + \partial_{r}F^{[rc]}$ with$ \partial_{r}(\nabla)^{2}F^{[rc]}= 0$. Let
$X^{c}:= \partial_{r}F^{[rc]};$ then, from  (\ref{cond1A})
$\partial_{c}(\partial_{a} X_{b} + \partial_{b} X_{a})= 0$. Whence follow the 
equation for a homothetic motion.

\subsection{Appendix 4}
If we use the notation $Z_1= T, Z_{i}= X_{i}, (i=1,2,3), Z_{j}= Y_{j}
(j=1,2,3)$ where $X_i$ correspond to time-dependent translations,
$Y_j$ to time-dependent rotations, the structure functions are given
by:\begin{eqnarray} c_{0i}^{~~i}= (ln f_{i})^{\dot{}} (i=1,2,3),
  c_{04}^{~~4}= (ln~\omega_{~3}^{2})^{\dot{}}, c_{05}^{~~5}= (ln~\omega_{~3}^{1})^{\dot{}},
  c_{06}^{~~6}=(ln~\omega_{~2}^{1})^{\dot{}} ~~~~~~~\nonumber\\ c_{ij}^{~~A}= 0~
  (i,j=1,2,3,~ A= 0, ..,6), ~c_{45}^{~~6}= -\frac{\omega_{~3}^{2}
    \omega_{~3}^{1}}{\omega_{~2}^{1}},~c_{56}^{~~4}= -\frac{\omega_{~3}^{1}
    \omega_{~2}^{1}}{\omega_{~2}^{3}}~c_{46}^{~~5}= -\frac{\omega_{~2}^{3}
    \omega_{~2}^{1}}{\omega_{~1}^{3}} \nonumber\\c_{14}^{~~A}= 0~ (A=
  0, ..,6),~c_{15}^{~~3}= -\frac{f_1}{f_3}\omega_{~3}^{1},~c_{16}^{~~2}=
  -\frac{f_1}{f_2}\omega_{~2}^{1}~~~~~~~~\nonumber\\c_{24}^{~~3}=
  -\frac{f_2}{f_3}\omega_{~3}^{2},~c_{25}^{~~A}= 0~ (A= 0, ..,6), ~c_{26}^{~~1}=
  \frac{f_2}{f_1}\omega_{~2}^{1},~~~~~~~~\nonumber\\c_{34}^{~~2}=
  \frac{f_3}{f_2}\omega_{~3}^{2},~c_{35}^{~~1}=\frac{f_3}{f_1}\omega_{~3}^{1},~c_{36}^
  {~~A}= 0~(A= 0, ..,6)~.~~~~~~~~~~ \nonumber \end{eqnarray}

\end{document}